\documentclass[,final,letterpaper]{aipproc}
\layoutstyle{8x11single}
\usepackage{amssymb}

\begin{document}

\title[Watch This Space]{Watch This Space: Observing Merging White Dwarfs}

\classification{97.80.-d,97.20.Rp,97.10.Cv,97.60.Bw}
\keywords{binary stars, white dwarfs, stellar evolution, supernovae}

\author{Ronald F. Webbink}{
address={Department of Astronomy, University of Illinois, 1002 W. Green St., Urbana, IL 61801, U.S.A.}}

\begin{abstract}

The Laser Interferometer Space Antenna (\emph{LISA}) will open the low-frequency (0.1-100 mHz) part of the gravitational wave spectrum to direct observation.  Of order 3600 galactic close binary white dwarfs will be individually resolvable in its all-sky spectrum, of which a dozen systems are expected to be on the verge of merger, showing the effects of strong tidal heating and/or early onset of tidal mass transfer.  Optical study of these systems would provide important insights into tidal dissipation mechanisms, and internal heating in merging white dwarfs that sets ignition conditions for potential type Ia supernovae.  Theoretical modeling and instrumentation programs are needed now to enable a campaign for optical identifications to exploit this opportunity.

\end{abstract}

\maketitle

\section{Introduction}

At this writing, the Laser Interferometer Space Antenna (\emph{LISA}) is targeted for launch in the mid-2020's.  This mission will open a new frequency window for gravitational wave detection, one which includes the expected signals of several thousand close double white dwarfs in the final stages of their inspiral toward merger.  Among these systems, some may have sufficient total mass to qualify as Type Ia supernova progenitors.  The purpose of this contribution is to call attention to the challenge and importance of identifying optically close double white dwarfs discovered by \emph{LISA} on the verge of merger.  

\emph{LISA} will consist of three spacecraft $5 \times 10^6\ {\rm km}$ apart, forming a roughly equilateral triangle, in what is basically a Michelson interferometer.  This formation of spacecraft will orbit the Sun in the ecliptic plane at $1\ {\rm AU}$, $20^{\circ}$ behind the Earth.  The three spacecraft themselves define a plane inclined by $60^{\circ}$ with respect to the ecliptic plane.  That plane precesses about the ecliptic pole as the spacecraft orbit the Sun, modulating the antenna pattern as they do so.  As \emph{LISA} orbits the Sun, periodic sources (such as double white dwarfs) will therefore be modulated in amplitude by this variation in antenna pattern, and in frequency by doppler motion of the spacecraft array with respect to the Sun.  These two effects combined provide a signature by which sources can be localized.

\emph{LISA} will be sensitive to GW frequencies in the range $\nu \approx 0.1$ to $100\ {\rm mHz}$. It is designed to achieve a limiting gravitational wave dimensionless strain amplitude of order $h \approx 10^{-23}$ at signal-to-noise ratio $SNR = 5$ in one year of operation \citep{princeetal02, rajeshnayaketal03}.  In the lower-frequency part of this window, galactic close double white dwarfs are expected to provide such a plethora of sources that \emph{LISA}'s sensitivity will effectively be confusion-limited by this background.  Only very nearby sources are likely to be detectable above this background.  However, as binary systems inspiral toward merger, the accelerating strength of GW losses leads to a characteristic power-law decline in the number of sources per unit frequency, $N_{\nu} \propto \nu^{-11/3}$.  When $\nu N_{\nu} \lesssim c/v_{\oplus} = 10^4$ ($v_{\oplus}$ being the Earth's orbital speed about the Sun), the average separation of sources in frequency becomes larger than their characteristic doppler width due to spacecraft orbital motion.  Individual sources are then resolvable.  \citet{webbinkhan98} estimate that this transition occurs around $\nu \approx 3.6\ {\rm MHz}$, and that of order $\sim 3600$ sources could then be resolved.  A more careful treatment of source confusion can be found in \citet{benderhils97}, yielding a similar estimate of the number of  GW sources resolvable in frequency.

\section{The Gravitational Wave Signature of Close Binary White Dwarfs}

The keys to identifying close double white dwarfs on the verge of merger are contained within their gravitational wave signatures.  Systems at this evolutionary stage are expected to have circular orbits, as universally observed among known close double white dwarf binaries.  The gravitational wave signal then immediately fixes the orbital period and phase of the source (modulo a $180^{\circ}$ ambiguity in phase because the GW frequency is twice the orbital frequency: $P_{\rm orb} = 4\pi/\nu$).  The dimensionless strain amplitude of a source a distance $d$ from the observer is \citep[cf.][]{petersmathews63, hilsetal90}
\begin{equation}
\label{ampl}
h = (h_+^2 + h_{\times}^2)^{1/2} = \left[ \frac{P_{\rm orb}^2}{2\pi^2 d^2 c^8} \left( \frac{2\pi Gm}{P_{\rm orb}} \right)^{10/3} \left( 1 + 6 \cos^2 i + \cos^4 i \right) \right]^{1/2} \ ,
\end{equation}
where $m$ is the ``chirp'' mass, a function of the binary component masses $M_1$ and $M_2$,
\begin{equation}
m \equiv \left( \frac{M_1^3 M_2^3}{M_1 + M_2} \right)^{1/5} \ ,
\end{equation}
and $h_+$ and $h_{\times}$ and the separate dimensionless amplitudes of the two polarizations of the (quadrupole) gravitational wave.  The ratio of $h_{\times}$ to $h_+$ is a function of the orbital inclination $i$ of the binary's orbital axis to the line of sight to the observer:
\begin{equation}
\label{pol}
\frac{h_{\times}}{h_+} = \frac{2 \cos i}{1 + \cos^2 i} \ .
\end{equation}
Thus, the gravitational wave signature of a source specifies not only the orbital ephemeris of that source, but, from the strain amplitude $h$ a measure of the functional combination $m^{5/3}/d$, and from the polarization a measure of the orbital inclination $i$.  In principle, the orientation of the gravitational wave polarization axes on the sky also informs us of the position angle of the orbital axis on the plane of the sky.  Models for the formation of close double white dwarfs suggest that chirp masses $m$ span only a relatively modest dynamical range: \citet{webbinkhan98} quote $\langle m \rangle = 0.42\ M_{\odot}$, with standard deviation $\sigma_m = 0.14\ M_{\odot}$ for model 4 in the population synthesis study by \citet{han98}.  The source amplitude $h$, combined with its polarization $h_{\times}/h_+$ enables one to make a very crude estimate of the source distance $d$.  Unfortunately, the gravitational wave signal alone does not permit extraction of the individual binary component masses $M_1$ and $M_2$.

Consider a fairly typical close double white dwarf of orbital period $P_{\rm orb} = 200\ {\rm s}$ and chirp mass $m = 0.42\ M_{\odot}$ at a distance  $d = 8\ {\rm kpc}$.  Averaging over random orbital inclinations, such a system will produce a dimensionless strain amplitude $h = 1.0 \times 10^{-22}$ at Earth, corresponding to a signal-to-noise ratio $SNR \approx 70$ at \emph{LISA}'s expected sensitivity at $\nu = 10\ {\rm mHz}$ \citep{princeetal02, rajeshnayaketal03}.  If subject to orbital energy losses due only to gravitational radiation, this source evolves toward higher frequency at a rate
\begin{equation}
\dot{\nu} = \frac{96\pi}{5} \left( \frac{\pi Gm}{c^3} \right)^{5/3} \nu^{11/3} = 6.3 \times 10^{-15} \ {\rm s^{-2}} \ .
\end{equation}
This rate of period change suffices to produce a phase drift of some 20 radians in one year, rapid enough to be readily detectable.  Measurement of $\dot{\nu}$ then permits determination of the chirp mass $m$ independent of the source distance.  The source distance itself can then be determined from the gravitational wave amplitude (see Eq.~\ref{ampl} above), given the orbital inclination from polarization (Eq.~\ref{pol}).  Indeed, it should be possible to measure $\dot{\nu}$ for the overwhelming majority of resolvable sources.

\section{The Approach to Mass Transfer}

Suppose that our typical binary consisted of equal-mass white dwarfs, $M_1 = M_2 = 2^{1/5}m = 0.48\ M_{\odot}$.  If the components of this binary were not subject to internal energy dissipation of any sort, but remained fully degenerate up to the onset of mass transfer, with orbital evolution driven purely by gravitational radiation, then that onset should occur after inspiraling for $t = 1.7 \times 10^4\ {\rm yr}$, when the orbital period had decayed to $P_{\rm orb} = 75\ {\rm s}$ ($\nu = 27\ {\rm mHz}$).  By that time the dimensionless strain amplitude $h$ should have grown by a factor of 1.9, and the drift rate of the gravitational wave frequency, $\dot{\nu}$, grown by a factor of 34.  

For comparison, consider also a system consisting of two $0.7\ M_{\odot}$ white dwarfs, again at a distance $d = 8\ {\rm kpc}$ with orbital period $P_{\rm orb} = 200\ {\rm s}$ and evolving solely under gravitational radiation.  Such a system would be of particular interest as a possible type Ia supernova progenitor.  It will produce a gravitational wave amplitude (again averaged over orbital inclination $i$) of $h = 1.9 \times 10^{-22}$ at $\nu = 10\ {\rm mHz}$, corresponding to $SNR \approx 130$.  The white dwarfs in this case are more compact than in the previous example, and their higher masses drive more rapid evolution.  After inspiraling for $t = 9.9 \times 10^3\ {\rm yr}$, this binary encounters the onset of mass transfer at orbital period $P_{\rm orb} = 43\ {\rm s}$.  The strain amplitude at that time reaches $h = 5.3 \times 10^{-22}$ ($SNR \approx 150$), and not only is $\dot{\nu} = 3.2 \times 10^{-12}\ {\rm s^{-2}}$ large enough to produce a phase shift of 3.9 rad in one week, but $\ddot{\nu} = 8.2 \times 10^{-22}\ {\rm s^{-3}}$ has grown large enough to produce a dramatic phase drift in its own right ($\Delta \phi = 54\ {\rm rad}$) over the course of one year's observation.

The approach to tidal mass transfer and merger in the two examples given above has been highly idealized.  In reality, additional dissipation mechanisms come into play, notably tidal heating and tidal mass transfer up to the point of runaway merger.  The issue of tidal heating on the approach to merger was raised by \citet{webbinkiben87}, and explored in considerably more detail by \citet{ibenetal98}.  It would seem almost certain that detached close double white dwarfs rotate asynchronously up to the point where mutual tidal perturbations become significant.  If we suppose them to rotate slowly, as isolated white dwarfs generally do, then tides must impart roughly 1\% of the orbital energy of the binary to rotation in order to bring the white dwarfs into synchronism.  This is not enough energy to lift degeneracy in the white dwarf interiors, but it is capable of substantially heating them \citep[cf.][]{ibenetal98}, thereby significantly influencing conditions under which carbon ignition may take place.  Nor is rotational energy large enough to alter qualitatively the course of orbital evolution; but the very short orbital evolutionary time scale prevailing immediately prior to merger, and the fact that tidal dissipation is strongest near the surface of the distorted star, imply that tidal heating can drive the dissipating star up to relatively high luminosities.

\subsection{Tidal Heating}

Iben et al. \citep{ibenetal98} have modeled tidal heating in a close binary white dwarf approaching merger, on the ad hoc assumption that the (spherically-averaged) local heating rate equals the local rate of increase in rotational energy per unit mass needed to maintain synchronous rotation.  They find that the surface luminosities of dissipative white dwarfs vary with orbital period as
\begin{equation}
L_2 \propto P_{\rm orb}^{-9} \ ,
\end{equation}
(see their Figure 6), with final luminosities at the onset of mass transfer approximately,
\begin{equation}
L_2 \sim 200\, (M/M_{\odot})^{6.1}\ L_{\odot} \ .
\end {equation}
These results suggest that white dwarf luminosities of order $2\ L_{\odot}$ to $20\ L_{\odot}$ (per white dwarf component) may be achieved immediately prior to tidal mass transfer in the first and second examples, respectively, posed above.  Characteristic growth time scales for these white dwarfs are of order $(L_2/\dot{L}_2) \approx -\frac{1}{9}(P_{\rm orb}/\dot{P}_{\rm orb}) = 430\ {\rm yr}\ \mbox{and}\  50\ {\rm yr}$, respectively.  These quantitative estimates of tidal heating depend entirely on the effective viscosity in the envelopes of the heated white dwarfs, and so are inherently uncertain.  Microscopic viscosity is negligible (as usual in astrophysical settings), but shear flows could generate enough turbulent dissipation, or amplify a seed magnetic field sufficiently, to drive synchronization on an inspiraling time scale.  Some significant degree of tidal heating would seem likely in any case.

\subsection{The Onset of Mass Transfer}

The onset of tidal mass transfer in an inspiraling system does not lead immediately to runaway mass transfer and merger, even when the binary satisfies the conditions for dynamical time scale mass transfer, but is preceded by a phase in which an initial trickle of mass is driven by orbital contraction, up to the point at which the mass transfer rate becomes comparable to the mass of the donor, divided by the gravitational wave inspiraling time scale.  With some idealizing assumptions (most notably, completely degenerate white dwarfs undergoing conservative mass transfer, apart from gravitational radiation losses), it is possible to solve analytically for the relation between mass transfer rate and time since contact, in the limit that the total mass transferred is small compared with the mass of the donor \citep{webbinkiben87}.

Let $\eta_2 \equiv (\partial \ln R_2/\partial t)_{M_2}$ be  the rate of expansion (or contraction) of the donor star in the absence of mass transfer ($\eta_2 = 0$ for cold white dwarfs); $\eta_L \equiv (\partial \ln R_L/\partial t)$ be the rate of change of the Roche lobe radius of the donor in the absence of mass transfer (in the present case, $\eta_L = -2/\tau_{\rm gr}$, where $\tau_{\rm gr}$ is the time scale characterizing  angular momentum loss by gravitational radiation); $\zeta_2 \equiv (\partial \ln R_2/\partial \ln M_2)_{\rm ad}$ be the adiabatic radius-mass exponent for the lobe-filling star; and $\zeta_L = (\partial \ln R_L/\partial \ln M_2)_t$ be the dependence of donor Roche lobe radius on donor mass in the presence of mass transfer, but absent any orbital angular momentum losses not coupled proportionally to the mass transfer rate.  The relation between time and mass transfer rate can than be written in dimensionless form:
\begin{equation}
t^* = - \frac{1}{2} \ln \left[ \frac{(1 - \lambda)^3}{(1 - \lambda^3)} \right] + \sqrt{3} \left[ \arctan \left( \frac{2\lambda + 1}{\sqrt{3}} \right) - \frac{\pi}{6} \right]\ ,
\end{equation}
where the dimensional time $t$ and mass transfer rate $\dot{M}_2$ are related to the dimensionless quantities $t^*$ and $\lambda$ through the relations $t = \tau t^*$ and $\dot{M}_2 = -\mu \lambda^3$.  The scale coefficients, $\tau$ for time and $\mu$ for mass transfer rate, are\footnote{We restore here a factor $P_{\rm orb}^{1/3}$ inadvertently omitted from the expression for $\tau$ in \citet{webbinkiben87}.}:
\begin{equation}
\tau = \frac{1}{3} \left[ F(q) \, E(M_2/M_{\rm Ch}) \, (\mu_e'/\mu_e)^{5/2} (\zeta_2 - \zeta_L) \right]^{-1/3} P_{\rm orb}^{1/3} \, ( \eta_2 - \eta_L)^{-2/3} \ ,
\end{equation}
\begin{equation}
\mu = M_2 \left( \frac{\nu_2 - \nu_L}{\eta_2 - \eta_L} \right) \ .
\end{equation}
Here, $\mu_e'$ is the electron mean molecular weight of the donor white dwarf envelope, and $\mu_e$ is the electron mean molecular weight of its interior.  Expressions for the dimensionless functions $F(q)$ (where $q = M_2/M_1$, the ratio of donor to accretor mass) and $E(M_2/M_{\rm Ch})$ (the ratio of donor white dwarf mass to the Chandrasekhar mass) are given by \citet{webbink84}.  Typically, they are of order $F \approx 0.015-0.03$, $E \approx 20-40$.

\begin{figure}[t]
%\begin{center}
\includegraphics[height=.3\textheight]{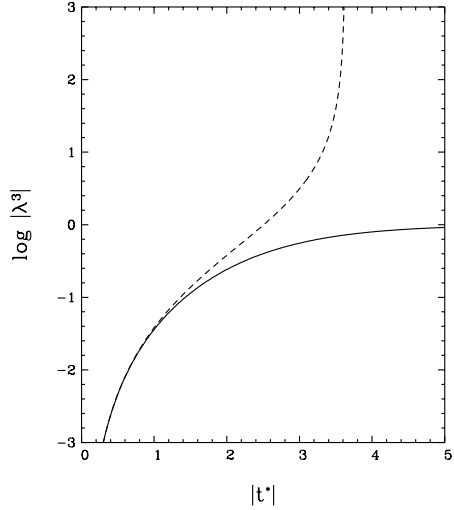}
\caption{Dimensionless mass transfer rate, $|\lambda^3|$, as a function of dimensionless time, $t^*$, for dynamically stable (solid line) and dynamically unstable (dashed line) mass transfer, for a fully-degenerate donor star.  Physical scaling factors for these two solutions are given in the text.}
\label{mdot-time}
%\end{center}
\end{figure}

There exist two branches to the solution for $t^*(\lambda)$, shown in Fig.~\ref{mdot-time}.  If $\zeta_2 > \zeta_L$, mass transfer is dynamically stable.  This situation holds only if the donor white dwarf is considerably less massive than the accretor, and never when $M_2 \ge 0.67\ M_{\odot}$, in the limit of conservative mass transfer. On this branch, $\tau$, $\mu$, $t^*$, and $\lambda$ are all positive.  The mass transfer rate asymptotically approaches $\dot{M}_2 = -\mu$ as $t \rightarrow \infty$. If  $\zeta_2 < \zeta_L$, mass transfer is dynamically unstable.  This is the situation for both examples here.  On this solution branch, $\tau$, $\mu$, $t^*$, and $\lambda$ are all negative.  The mass transfer rate becomes singular at $t^* = 2\pi/\sqrt{3} $.

For uniform composition and equal component masses, the scaling coefficients in the two examples give above are $\tau = -1.4\ {\rm yr}$ (time from onset of mass transfer to merger $t = -(2\pi/\sqrt{3})\tau = 5\ {\rm yr}$), $\mu = -1.3 \times 10^{-4}\ M_{\odot}\,{\rm yr^{-1}}$ for the typical close double white dwarf (assuming equal component masses), and $\tau = -0.29\ {\rm yr}$  ($t = 1.1\ {\rm yr}$), $\mu = -8.3 \times 10^{-4}\ M_{\odot}\,{\rm yr^{-1}}$ for the putative type Ia supernova progenitor.  In the examples at hand, the binary components fill their Roche lobes simultaneously in both cases, so there is no difference in gravitational potential between the binary components.  But for even slightly different component masses, these characteristic mass transfer rates would drive the accretion luminosities up to their respective Eddington limits.

\section{The Population of Close Double White Dwarfs on the Verge of Merger}

It is clear from the above discussion that, far from lurking in anonymity at low luminosity up to the instant of merger, close double white dwarfs can be expected to advertise their imminent demise by gradually brightening over the decades or centuries leading up to merger, first by tidal dissipation in the envelopes of their donor (less massive) components, then by a dramatic rise in accretion luminosity over the final year to decade preceding merger\footnote{In reality, the duration of this latter, early mass transfer, phase has been underestimated here.  The outer envelope of the donor white dwarf will not be electron-degenerate, as assumed above, but relatively more extended, softening the onset of tidal mass transfer.}.  

The obvious question is whether the event rates for close binary white dwarf mergers are high enough to be observable.  If we adopt again the population synthesis model \citep{han98} cited above, it predicts a total galactic white dwarf merger rate of $0.030\ {\rm yr^{-1}}$.  Given the estimates obtained above for the characteristic tidal heating and mass transfer time scales of a typical close double white dwarf system, of order a dozen systems in the Galaxy should lie within their final $e$-folding tidal heating time scale before merger, with a probability of $\ge 20\%$ that at least one of these systems has begun tidal mass transfer.  The same population synthesis model yields a galactic white dwarf merger rate for super-Chandrasekhar mass systems of $3.6 \times 10^{-3}\ {\rm yr^{-1}}$, consistent with estimates of the galactic SNIa rate.  There is about a 20\% probability that one of these systems galaxy-wide will lie within its final $e$-folding tidal heating time scale before merger, but only a $\sim 0.4\%$ probability that a super-Chandrasekhar mass system has begun tidal mass transfer and will merge within the next year.

\section{Optical Identification of \emph{LISA} Sources}

There are important astrophysical issues that could be addressed by optical observations of these close binary white dwarfs in their final approach to merger.  The magnitude and efficiency of tidal heating is of course one of them.  Do the component white dwarfs ever approach synchronism?  To what extent are the white dwarf interiors reheated by tidal dissipation?  The calculations published by Iben et al. \citep{ibenetal98} indicate that such reheating could be substantial, and could be exercise major influence on conditions under which carbon ignition occurs in merging CO white dwarfs.  Gravitational wave observations alone provide only limited insight into this question: if $\ddot{\nu}$ can be measured, in addition to $\nu$ and $\dot{\nu}$, then its departure from the case for pure gravitational radiation, $\ddot{\nu} = \frac{11}{3} \dot{\nu}^2/\nu$, must be laid to orbital energy losses (such as tidal heating,  mass transfer, or mass loss) other than gravitational radiation.  Of great importance would be determination of the mass ratios of these incipiently merging close double white dwarfs.  Gravitational wave observations alone constrain only the chirp mass $m$, while the orbital period at the onset of mass transfer is dictated essentially by the mean density (and hence by the mass) of the donor white dwarf.  Absent knowledge of the binary mass ratio, the dispersion in mass ratios at a given chirp mass translates into a large uncertainty in the time-to-merger of a given system, even late in its evolution.  But with a mass ratio in hand (and certainly in the case where the secular growth in system luminosity is measurable), it becomes possible to reduce the uncertainty in time-to-merger to a human time scale, possibly to decades or even years.  The chance is remote, but nonzero, that it will be possible not only to identify the next type Ia supernova, but to predict the date of its explosion (assuming prompt detonation on merger) within a year or so!

The identification of optical counterparts to \emph{LISA} sources presents a major challenge.  \emph{LISA}'s angular resolution is poor, of order
\begin{equation}
\Delta \Omega \sim \frac{7.5}{1+ 0.8(\nu/{\rm mHz})^2} \frac{1}{SNR^2}\ {\rm sr}
\end{equation}
\citep[cf.][]{cutler98}.  Sources at $\nu = 10\ {\rm mHz}$, say, will require search boxes of several square degrees; and despite the relatively high luminosities systems on the verge of merger may reach, they are likely to remain very compact and hot, with only modest visual luminosities.  Furthermore, if these systems share the same scale height as old stars in the galactic disk, those at distances exceeding one or a two kpc are likely to be heavily reddened, and observable only in the near infrared.

The key to identifying close double white dwarfs will be the fact that their ephemerides are known from their \emph{LISA} discovery observations, and that, notwithstanding their astrophysically rapid inspiral time scales, their orbital periods still remain are stable enough in the short run to be useful, even if the binary phase drifts significantly.  These systems will be anisotropic emitters, ellipsoidal variables at the very least, and possibly much more strongly modulated if caught during the onset of mass transfer.  Their light will be modulated with the orbital period.  The key to identification is then to accumulate images of the candidate field of view phased with the binary period.  That period in the most interesting cases will be so short that it is impractical to read out a large-format CCD several times per binary orbit.  If the field of view is sufficiently uncrowded, the image can be dithered with the orbital period; if it is crowded, however, one may need to separate images into distinct phase bins, for example by chopping the image synchronously with the binary phase into separate, non-overlapping images.

\section{Conclusions}

The purpose of this contribution is to stimulate the groundwork for fully exploiting the expected return from \emph{LISA}.  More realistic theoretical models are needed for tidal heating (including seed magnetic fields) and for the onset of mass transfer, to provide templates for the observational signatures of systems on the verge of merger.  On the observational side, instrumental capability needs to be developed to accumulate rapid, phased observations of large fields of view.

\begin{theacknowledgments}

I want to thank Vicky Kalogera for spearheading this wonderful conference, the Local Organizing Committee for arranging such an enchanting venue, and the other members of the Scientific Organizing Committee for such a stimulating agenda.  I also thank Peter Bender and Dieter Hils for introducing me to \emph{LISA} when it was still a gleam in their eyes, Zhanwen Han for providing details of his population synthesis studies, and Matt Benacquista and Tony Tyson for useful discussions of \emph{LISA} and \emph{LSST} capabilities.  This research was supported in part by NSF grants 9618462 and 0406726.  Participation in this conference was supported by a grant from the Department of Astronomy, University of Illinois at Urbana-Champaign.

\end{theacknowledgments}


\begin{thebibliography}{11}

\bibitem[Bender and Hils(1997)]{benderhils97}P.~L. Bender and D. Hils, \emph{Class. Quantum Grav.}, \textbf{14}, 1439--1444 (1997).

\bibitem[Cutler(1998)]{cutler98}C. Cutler, \emph{Phys. Rev. D}, \textbf{57}, 7089--7102 (1998).

\bibitem[Han(1998)]{han98}Z. Han, \emph{Monthly Notices R. Astr. Soc.}, \textbf{296}, 1019--1040 (1998).

\bibitem[Hils, Bender and Webbink(1990)]{hilsetal90}D. Hils, P.~L. Bender, and R.~F. Webbink \emph{Astrophys. J.}, \textbf{360}, 75--94 (1990)  [Erratum: \emph{Astrophys. J.}, \textbf{369}, 271 (1991)].

\bibitem[Iben, Tutukov and Fedorova(1998)]{ibenetal98}I. Iben, Jr., A.~V. Tutukov, and A.~V. Fedorova \emph{Astrophys. J.}, \textbf{503}, 344--349 (1998).

\bibitem[Peters and Mathews(1963)]{petersmathews63}P.~C. Peters and J. Mathews, \emph{Phys. Rev.}, \textbf{131}, 435--440 (1963).

\bibitem[Prince et al.(2002)]{princeetal02}T.~A. Prince, M. Tinto, S.~L. Larson, and J.~W. Armstrong, \emph{Phys. Rev. D}, \textbf{66}, 122002 (2002).

\bibitem[Rajesh Nayak et al.(2003)]{rajeshnayaketal03}K. Rajesh Nayak, A. Pal, S.~V. Dhurandhar, and J.-Y. Vinet, \emph{Class. Quantum Grav.}, \textbf{20}, 1217--1231 (2003).

\bibitem[Webbink(1984)]{webbink84}R.~F. Webbink \emph{Astrophys. J.}, \textbf{277}, 355--360 (1984).

\bibitem[Webbink and Han(1998)]{webbinkhan98}R.~F. Webbink and Z. Han, ``Gravitational Radiation from Close Double White Dwarfs,'' in \emph{Laser Interferometer Space Antenna}, edited by W.~M. Folkner, AIP Conference Proceedings 456, American Institute of Physics, New York, 1998, pp. 61--67.

\bibitem[Webbink and Iben(1987)]{webbinkiben87}R.~F. Webbink and I. Iben, Jr. ``Tidal Interaction and Coalescence of Close Binary White Dwarfs,'' in \emph{The Second Conference on Faint Blue Stars, I.A.U. Colloq. No. 95}, edited by  A.~G.~D. Philip, D.~S. Hayes, and J.~W. Liebert, L. Davis Press, Schenectady, 1987, pp. 445--455.

\end{thebibliography}
\end{document}